# Resurrection of Traditional Luminosity Evolution Models to Explain Faint Field Galaxies[1]

Caryl Gronwall and David C. Koo
University of California Observatories / Lick Observatory,
Board of Studies in Astronomy and Astrophysics,
University of California, Santa Cruz, California 95064

## ABSTRACT

We explore the nature of the evolution of faint field galaxies by assuming that the local luminosity function is *not* well defined. We use a non-negative least squares technique to derive a near optimal set of *local* luminosity functions for different spectral types of galaxies by fitting to the observed optical and near-infrared counts, $B - R$ colors, and redshift distributions for galaxies with $15 \leq B \leq 27$. Our previous work showed that a no-evolution model for the luminosity functions was able to match the observed blue galaxy counts to within a factor of less than 50% by $B \sim 25$ versus the 5× to 15× (*e.g.*, Tyson 1988) of other non-evolving models. We report here the results of using only traditional luminosity evolution (*i.e.*, the photometric evolution of stars in a galaxy over time given reasonable assumptions of the form of the star formation history for various galaxy types), and find excellent fits to the observed data to $B \sim 23$. The addition of simple reddening with an SMC extinction law to our model spectral energy distributions extends the almost perfect fits to the faintest limits. While prior luminosity evolution models required both a low $q_0$ and a high galaxy formation redshift to fit the observed data, the quality of our fits is not significantly degraded by changing the value of $q_0$ to 0.5. We conclude that models more exotic than traditional luminosity evolution are not yet *required* to explain existing faint field galaxy data and thus the need for contributions by mergers or new populations of galaxies is at least 5× less than previously estimated (*e.g.*, Broadhurst *et al.* 1992).

*Subject headings:* galaxies: evolution — galaxies: luminosity function, mass function — galaxies: photometry — cosmology: observations

---

## 1. INTRODUCTION

Many workers have concluded that there is an excess of observed galaxy counts above non-evolutionary (NE) models of anywhere from a factor of two by $B \sim 20$ or $\sim 22.5$, to factors of 5 to 15 by $B \sim 25$. These blue counts have been claimed, even with evolution, to be incompatible with a flat ($\Omega = 1$) Friedmann universe, or with the observed near-infrared K band counts. Recent redshift surveys, however, have shown that galaxies with $B > 20$ have redshift distributions close to those predicted by NE models. These results have led to the introduction of many exotic theories such as rapid merging, disappearing populations of dwarf galaxies, heuristic forms of luminosity evoution, or the adoption of a cosmological constant. A discussion of and references for these and other claims can be found in Lilly (1993) and Koo, Gronwall, & Bruzual (1993, hereafter KGB).

Koo & Kron (1992) suggested, however, that the faint field galaxy data could be explained without exotic theories of evolution. They found that a NE model (generated by *trial and error*) fit the data moderately well. Improving on this approach, KGB assumed that the local luminosity function was not well-defined and used a non-negative least squares technique to find the best fitting NE model. This model, which includes a steep faint end slope to the luminosity function (see also Driver *et al.* 1994), fits the data extraordinarily well. Despite the vastly improved fits of the KGB NE model, we expect that real galaxies undergo some evolution as they age. Moreover, there were sufficient discrepancies between the data and the NE model to indicate that some luminosity evolution might improve the quality of the fits. Thus, in this *Letter*, we use the method of KGB, but include traditional luminosity evolution (*i.e.*, the photometric evolution of stars in a galaxy over time given reasonable assumptions of the form of the star formation history for various galaxy types) and simple galaxy reddening to produce a "best-fitting" *local* luminosity function.

## 2. OBSERVATIONS AND METHOD

Our method for determining the galaxy luminosity functions is described by KGB. We find the weights to be applied to a set of input basis vectors that best fit (using non-negative least squares) the observations. These vectors in our current analysis are the predicted number distributions of colors and redshifts for 154 combinations of 14 galaxy absolute luminosity distributions and 11 SED types (see Table 1 and KGB for a complete description). The output weights for these 154 kinds of galaxies are proportional to their volume densities, and thus provide a direct measure of the *local* LF for each SED type.

An open universe with $q_0 = 0.05$ and a Hubble constant of 50 km-sec$^{-1}$-Mpc$^{-1}$ is adopted for the model predictions. Changing the value of $q_0$ to 0.5, however, has little effect on the quality of the model fits to the data. The SEDs were created with the star formation histories specified in Table 1 and with traditional luminosity evolution using the models of Bruzual & Charlot (1993). We assume that the bluest galaxy types (our non-evolving, constant SFR models at different ages) are due to galaxies undergoing bursts of star formation and that the number of galaxies per unit co-moving volume in a particular blue SED state is constant with time without making any assumptions about how individual objects evolve (see Koo 1985). By choosing an old age for galaxy formation ($z_f \geq 5$) and an old universe ($h = 0.5$), the effects of luminosity evolution are relatively mild. We reddened (see also Wang 1991) these model SEDs (for all but the two reddest SEDs) using an SMC extinction law (Bouchet et al. 1985) and $E(B - V) = 0.1$. Kunth & Mass-Hesse (1993) have found that the reddening in blue star-forming galaxies follows an SMC extinction law regardless of metallicity. This simplistic model of reddening in external galaxies is intended to provide a good "on average" estimate of the effects of dust in faint galaxies.

The data to be fit are organized as a vector matched to the model input vectors and include observed galaxy $B_J - R_F$ colors and redshift distributions for different $B_J$ magnitude intervals as well as $B_J$, Gunn $r$, and near-IR $K$-band counts. The bulk of these data comes from the sources listed in Koo & Kron (1992) and KGB. We have also included $B - V$ colors for bright galaxies provided to us by Kron & Takamiya (1994) based on unpublished observations of Zwicky galaxies taken by G.E. Kron and C.D. Shane in the 1960's. All of these surveys have been renormalized to agree with the observed $B_J$ galaxy counts including the most recent very faint CCD counts from Metcalfe et al. (1994b). We also correct for the systematic errors in bright galaxy counts discovered by Metcalfe, Fong, & Shanks (1994a). We fit the above data to a limiting magnitude of $B_J = 24$ for the redshifts and $B_J = 27$ for the colors. Observational errors in color were included by an appropriate Gaussian smoothing of the model data. The results also depend on the relative weighting adopted for different parts of the observational data set. As a default, weights are proportional to the observed number of galaxies in each bin. We have chosen, however, to give more emphasis to the redshift data and bright color data by weighting by five times the observed numbers in each bin. We also fit to the number of galaxies per 0.1 square degree rather than per 1 square degree to prevent the domination of the large galaxy counts over the other data.

### 3. RESULTS AND DISCUSSION

Table 1 gives the best-fitting output LFs, as observed locally, for each SED type using the reddened SEDs. The color-integrated LFs of our fit are compared to recent observational derivations in Figure 1. Our total LF is in good agreement with the flat local LF measured by Loveday et al. (1992) to $M_{B_J} \sim -17$. Our predicted LF does, however, show a rise at fainter magnitudes. This rise occurs at magnitudes fainter than those actually measured by Loveday et al. and is fully compatible with the LFs derived from fainter redshift surveys (Eales 1993; Lonsdale & Chokshi 1993) as well as with the local LF measurements of Marzke *et al.* (1994b). Subdivided by color, our LFs agree over the range measured with those of of Metcalfe *et al.* (1991) and Marzke *et al.* (1994b).

The galaxy counts, colors, and redshift distributions predicted for our derived set of local LFs are displayed in Figures 2-4 along with the observations and the predictions of a best-fitting no-evolution model (an update of the KGB model using the most recent data described above). The fits of the evolutionary model to the data are excellent. The fit to the $B_J$ counts, shows a less than 10% deficit compared to the observed counts at $B_J = 25$. The fit to the red (Gunn $r$) counts is also good and is consistent with the most recent $r$ counts of Weir (1994) which lie slightly below the average. (Note that there was an error in the $r$ band zeropoint in the KGB model; their model curve should be shifted by 0.43 mag towards fainter magnitudes.) We find slight discrepancies at the faintest $K$ magnitudes which may be caused by uncertainties in the faintest counts. The predicted color distributions agree quite well with the observations up to $B_J = 27$. This model also matches the faint galaxy colors of Steidel & Hamilton (1993). When reddening was not included in the input SEDs, the observed colors were fit well only to $B_J = 23$, while at fainter magnitudes the predicted colors were $\sim 0.3$ to 0.4 magnitudes bluer than observed. The redshift distributions predicted by our model match the observations well to $B_J = 23$ (see Figure 4). For $B_J > 23$ the model predicts a significant fraction ($\sim 20\%$) of high redshift ($z > 1$) galaxies which have not been observed in existing redshift surveys. This discrepancy may suggest some new evolutionary effect, but may instead be due to observational biases and incompleteness or to uncertainties in our models.

In conclusion, using only traditional luminosity evolution and simple galaxy reddening, we are able to fit well the observed optical and near-infrared counts, $B - R$ colors, and redshift distributions for galaxies with $15 \leq B \leq 27$. We find essentially no excess of blue galaxy counts above our best-fitting model. The quality of the fits is not significantly degraded by a change in $q_0$ to 0.5. Previous workers were able to fit the observed data only by invoking heuristic models for the luminosity evolution (*e.g.*, Lilly 1993; Broadhurst *et al.* 1988) or by using traditional luminosity evolution with a low $q_0$ and a large galaxy formation redshift (e.g., Yoshii & Peterson 1991, Guiderdoni & Rocca-Volmerange 1990). Both the data and models have uncertainties and improved fits can be expected in the future. Nonetheless, given the excellent quality of our fits and without claiming uniqueness, we conclude that the existing faint field galaxy data can be almost totally explained by a traditional luminosity evolution model and that the *need* for mergers or new populations of galaxies is at least 5× smaller than claimed by others (*e.g.*, Broadhurst *et al.* 1992).

We thank G. Bruzual for providing us with his galaxy evolution software, the referee for useful suggestions, B. McLeod for pointing out our $r$-band zeropoint error, and N. Weir, C. Steidel, R. Kron, M. Takamiya, and R. Marzke for providing their observational data. D. C. K. gratefully acknowledges support for this work from an US-Venezuela NSF grant INT-9003157, an NSF PYI grant AST-8858203, and two faculty research grants from UC Santa Cruz. C. G. acknowledges support from an NSF Graduate Fellowship.

# Figure Captions

FIG. 1a. – Derived luminosity functions versus previously derived ones, all scaled to $H_0 = 50$ km-sec$^{-1}$-Mpc$^{-1}$. Thick-lined histogram is the derived total differential luminosity function; thin line is from the derived Schechter LF of Loveday *et al.* (1992) and is valid for $M_{B_J} \leq -16.75$ (dashed line indicates extrapolation of this fit). Circles are data points from Loveday *et al.* (1992), squares from Marzke *et al.* (1994) (arbitrarily normalized to agree at $M_{B_J} = -18.5$) plus signs from Eales (1993), and stars from Lonsdale & Chokshi (1993). For clarity, we have not included the authors' original error bars on these points.

FIG. 1b. – Same as for 1a, except thick-lined histogram applies for color classes with $B - V \geq 0.85$. The thin line is the differential LF from Table 8 of Metcalfe *et al.* (1991) for the same color range and valid for $M_{B_J} \leq -17.5$; dashed line indicates extrapolation of their fit.

FIG. 1c. – Same as for 1b except for $0.6 < B - V < 0.85$ and thin line being valid for $M_{B_J} \leq -15.5$; dashed line indicates extrapolation of their fit.

FIG. 1d. – Same as for 1c except for $B - V \leq 0.6$.

FIG. 2. – Log of the differential counts A (per mag per square degree) of faint field galaxies versus magnitude in the indicated bands. $B_J$ and Gunn $r$ band observations (open squares and solid circles) are compilations of data made by Koo & Kron (1992); the infrared $K$ band counts (large open circles) are a compilation from Gardner *et al.* (1993). Solid squares show the bright $B_J$ counts corrected for the systematic errors discovered by Metcalfe *et al.* (1994a) as well as the new faint CCD counts of Metcalfe *et al.* (1994b); small open circles show the new $r$ band counts of Weir (1994). Our new mild evolution (with reddening) predictions are shown as solid curves while the equivalent best-fitting no-evolution predictions are shown as dotted curves.

FIG. 3. – Color distributions versus indicated magnitude intervals are shown as thick lines for the observations and thin solid lines for the mild evolution model predictions and dotted lines for the NE predictions. The $B = 14 - 16$ interval shows $B - V$ color with the observations provided by Kron & Takamiya (1994). Other panels show $B_J - R_F$ color distributions in the given $B_J$ magnitude intervals with the observations from the data of several groups compiled by (Koo & Kron 1992).

FIG. 4. — Normalized histograms versus redshift ($\log z$) for magnitude ($B_J$) intervals as indicated. Bin size is 0.143, since original bin size was 1.0 in 7 $\log z$. The evolutionary model predictions are shown as thin solid lines while the NE model predictions are shown as dotted lines; observations shown as thick lines are compilations from different surveys (Koo & Kron 1992); the $B_J = 23$ to 24 observations are updated from Colless *et al.* (1993a,b).

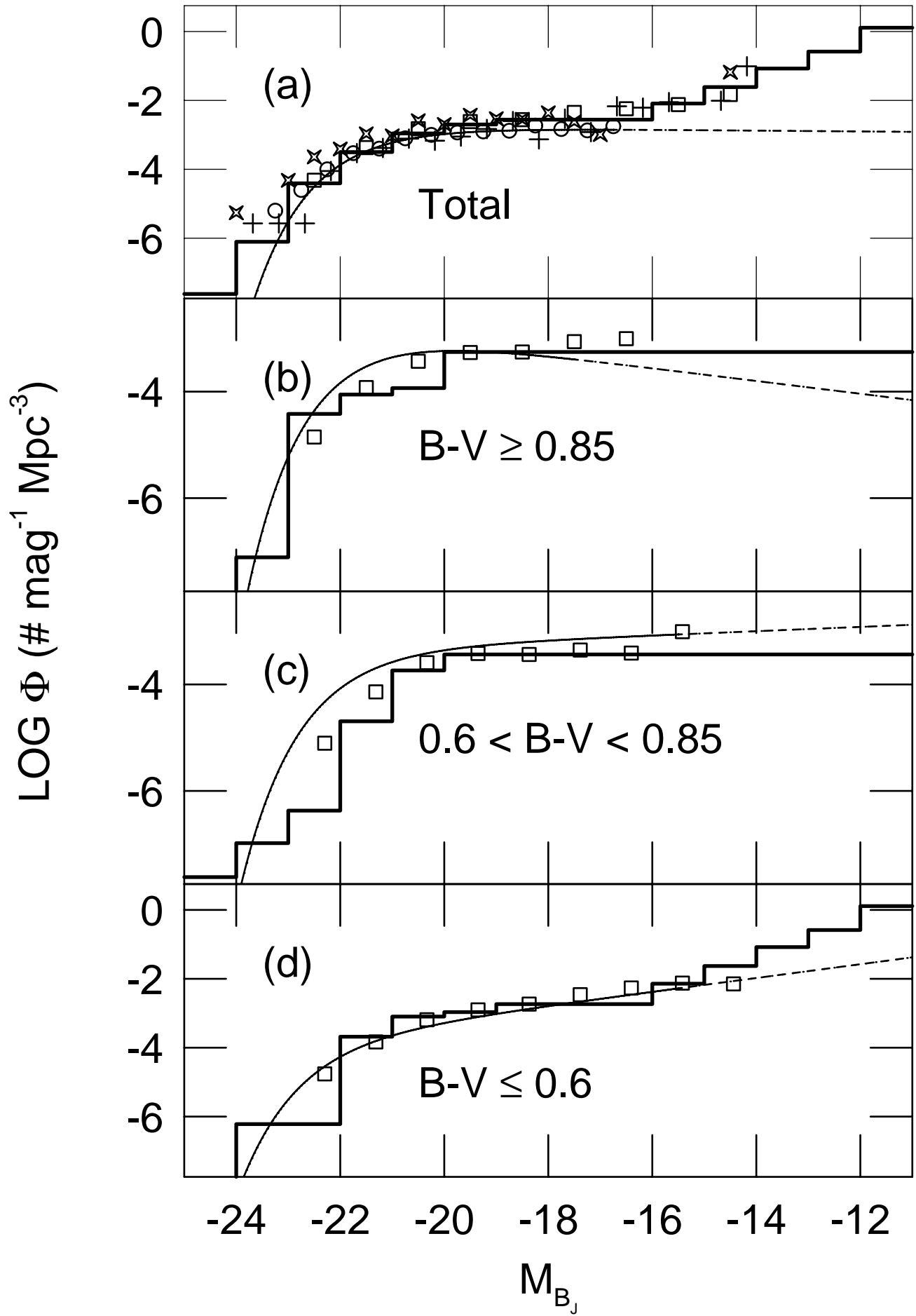

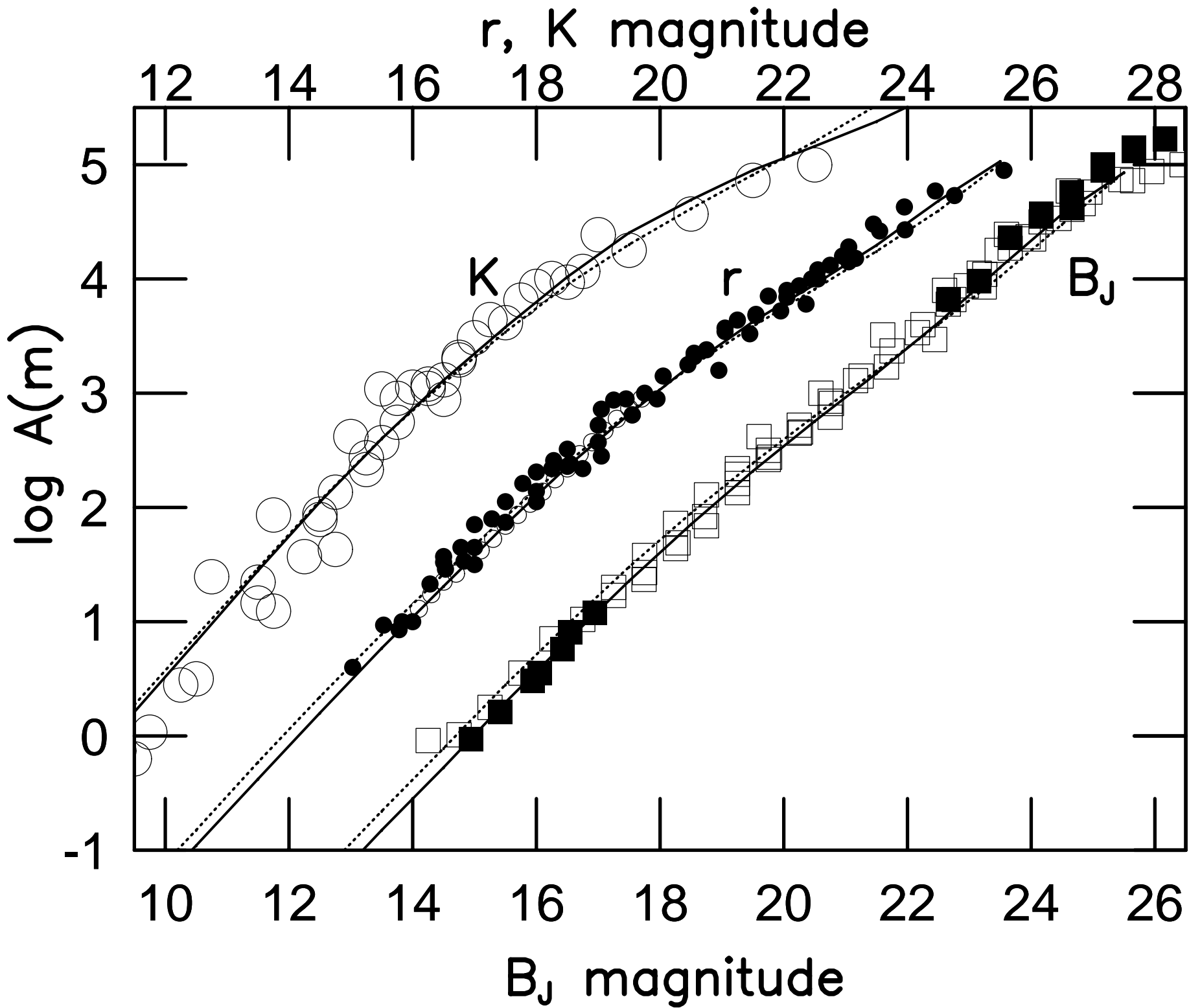

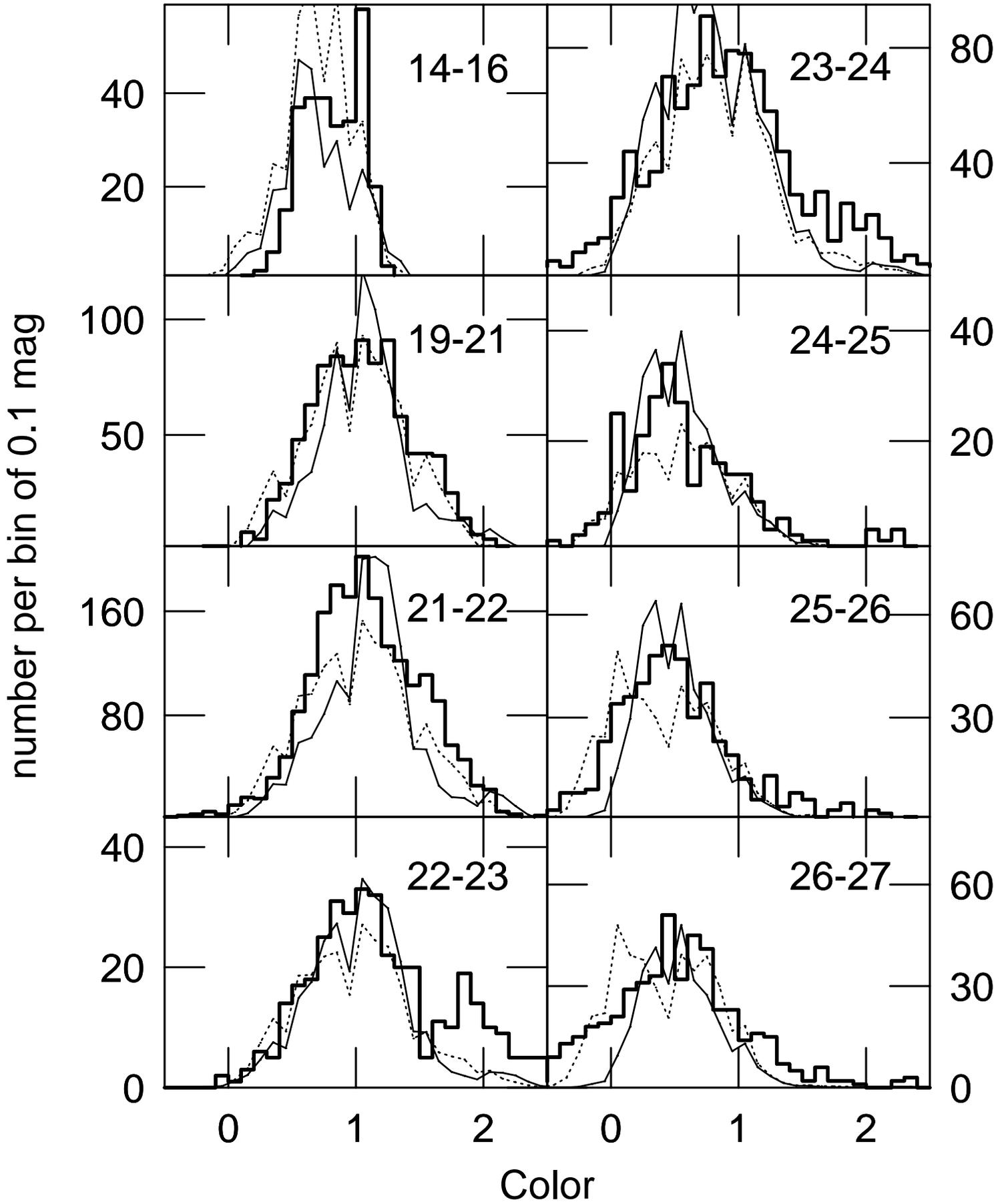

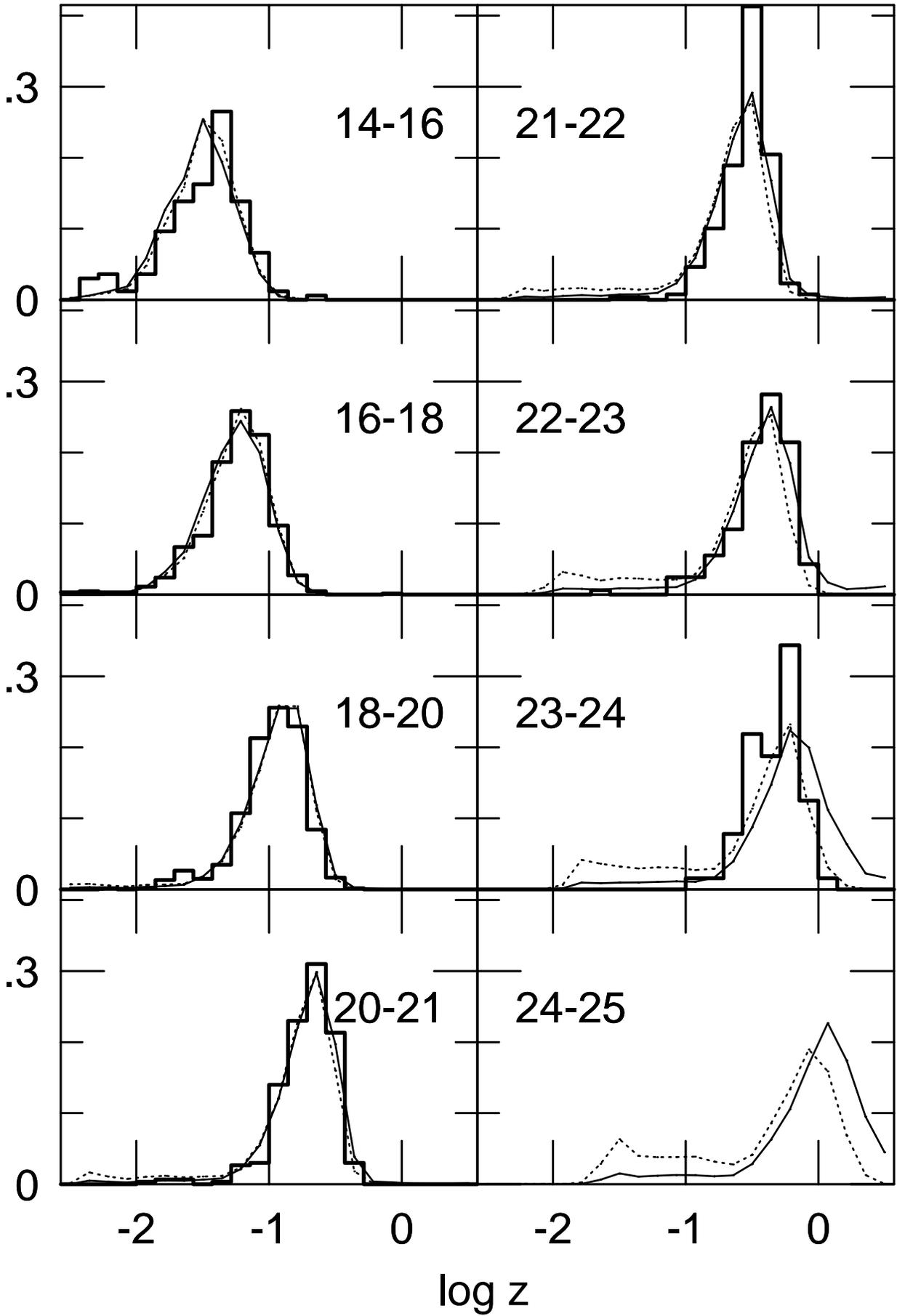

TABLE 1
INPUT GALAXY COLOR CLASSES AND
OUTPUT MODEL LUMINOSITY FUNCTIONS[a]

| Class | 1 | 2 | 3 | 4 | 5 | 6 | 7 | 8 | 9 | 10 | 11 | | |
|---|---|---|---|---|---|---|---|---|---|---|---|---|---|
| Age[b] | 0.4 | 2.0 | 6.8 | 16.0 | 16.0 | 16.0 | 16.0 | 16.0 | 16.0 | 16.0 | 20.0 | | |
| $\mu$[c] | C | C | C | 0.01 | 0.10 | 0.15 | 0.20 | 0.25 | 0.30 | 0.70 | B | | |
| $B-V$ | 0.15 | 0.25 | 0.35 | 0.43 | 0.52 | 0.61 | 0.70 | 0.78 | 0.85 | 0.95 | 0.99 | | |
| $B-V$[d] | 0.24 | 0.34 | 0.44 | 0.51 | 0.59 | 0.67 | 0.75 | 0.83 | 0.90 | 0.95 | 0.99 | | |
| $M_{B_J}$[e] | | | | | | | | | | | | Total | Obs[f] |
| $-24.5$ | ... | ... | ... | ... | ... | ... | $-7.62$ | ... | ... | ... | ... | $-7.62$ | $-10.8$ |
| $-23.5$ | ... | ... | ... | ... | $-6.22$ | ... | $-7.14$ | $-7.48$ | ... | $-7.11$ | ... | $-6.10$ | $-6.32$ |
| $-22.5$ | ... | ... | ... | ... | $-6.22$ | ... | $-6.40$ | $-7.48$ | $-6.57$ | $-6.15$ | $-4.43$ | $-4.41$ | $-4.35$ |
| $-21.5$ | ... | ... | ... | $-3.78$ | $-4.38$ | ... | $-4.83$ | $-5.25$ | $-6.57$ | $-5.24$ | $-4.08$ | $-3.50$ | $-3.46$ |
| $-20.5$ | $-4.43$ | ... | ... | $-3.78$ | $-3.22$ | ... | $-4.83$ | $-3.77$ | $-4.67$ | $-4.88$ | $-4.08$ | $-2.96$ | $-3.07$ |
| $-19.5$ | $-3.50$ | ... | ... | $-3.78$ | $-3.22$ | $-5.22$ | $-3.72$ | $-3.77$ | $-3.42$ | $-3.88$ | $-4.08$ | $-2.70$ | $-2.92$ |
| $-18.5$ | $-3.05$ | $-3.94$ | ... | $-3.78$ | $-3.18$ | $-5.22$ | $-3.72$ | $-3.77$ | $-3.42$ | $-3.88$ | $-4.08$ | $-2.56$ | $-2.87$ |
| $-17.5$ | $-3.05$ | $-3.94$ | ... | $-3.78$ | $-3.18$ | $-5.22$ | $-3.72$ | $-3.77$ | $-3.42$ | $-3.88$ | $-4.08$ | $-2.56$ | $-2.85$ |
| $-16.5$ | $-3.05$ | $-3.94$ | ... | $-3.78$ | $-3.18$ | $-5.22$ | $-3.72$ | $-3.77$ | $-3.42$ | $-3.88$ | $-4.08$ | $-2.56$ | $-2.85$[g] |
| $-15.5$ | $-3.05$ | $-3.94$ | ... | $-3.78$ | $-3.18$ | $-5.22$ | $-3.72$ | $-3.77$ | $-3.42$ | $-3.88$ | $-4.08$ | $-2.09$ | $-2.86$[g] |
| $-14.5$ | $-2.62$ | $-3.94$ | ... | $-2.26$ | $-1.81$ | $-5.22$ | $-3.72$ | $-3.77$ | $-3.42$ | $-3.88$ | $-4.08$ | $-1.61$ | $-2.87$[g] |
| $-13.5$ | $-2.62$ | $-3.94$ | ... | $-2.26$ | $-1.12$ | $-5.22$ | $-3.72$ | $-3.77$ | $-3.42$ | $-3.88$ | $-4.08$ | $-1.07$ | $-2.88$[g] |
| $-12.5$ | $-2.62$ | $-3.94$ | ... | $-2.26$ | $-0.60$ | $-5.22$ | $-3.72$ | $-3.77$ | $-3.42$ | $-3.88$ | $-4.08$ | $-0.58$ | $-2.89$[g] |
| $-11.5$ | $-0.15$ | $-3.94$ | ... | $-2.26$ | $-0.24$ | $-5.22$ | $-3.72$ | $-3.77$ | $-3.42$ | $-3.88$ | $-4.08$ | $+0.11$ | $-2.91$[g] |

[a] $\log \Phi$ (Number-mag$^{-1}$-Mpc$^{-3}$); $H_0 = 50$ km-s$^{-1}$-Mpc$^{-1}$; ...indicates no galaxies in bin
[b] Age of the galaxy color class in Gyr.
[c] Star formation history of the galaxy class: All color classes have Salpeter initial mass functions for masses between 0.1 and $125 M_\odot$. Classes 1-3 have a constant (C) star formation rate with no evolution in the SED. Classes 4-10 have an exponentially decreasing star formation rate parameterized by $\mu$, where $\mu$ is the fraction of mass converted into stars in 1 Gyr. Class 11 is a $10^7$ Gyr burst (B) of star formation.
[d] $B - V$ color with extinction of $E(B - V) = 0.1$ and an SMC extinction law. Classes 10 & 11 have no reddening.
[e] Center of one magnitude bins
[f] Luminosity function from Loveday et al. (1992): $M^* = -21$, $\alpha = -0.97$, $\Phi^* = 1.75 \times 10^{-3}$
[g] Loveday et al. luminosity function only valid for $M_{B_J} \leq -16.75$; these are extrapolations of the $\alpha = -0.97$ slope